# The Structure of Eu-III

**R J Husband[1], I Loa[1], G W Stinton[1], S R Evans[2], G J Ackland[1] and M I McMahon[1]**

[1]SUPA, School of Physics and Astronomy and Centre for Science at Extreme Conditions, The University of Edinburgh, Edinburgh, EH9 3JZ, UK
[2]European Synchrotron Radiation Facility, 38043 Grenoble, France

E-mail: R.J.Husband@sms.ed.ac.uk

**Abstract.** Previous x-ray diffraction studies have reported Eu to transform from the *hcp* structure to a new phase, Eu-III, at 18 GPa. Using x-ray powder diffraction we have determined that Eu remains *hcp* up to 33 GPa, and that the extra peaks that appear at 18 GPa are from an impurity phase with space group $R\bar{3}c$. Above 33 GPa the diffraction pattern becomes very much more complex, signalling a transition to a phase with a distorted *hcp* structure.

## 1. Introduction

The majority of the lanthanide elements are trivalent due to their $4f^n 5d^1 6s^2$ outer electronic structure. The two exceptions are europium (Eu) and ytterbium (Yb), which are divalent due to their half-filled and filled 4f shells, respectively. As a result, Eu and Yb both exhibit a significantly larger atomic volume than would be expected to be consistent with the general trend observed within the lanthanide series [1]. They also do not follow the general trend of phase transitions under pressure or with increasing atomic number observed in the trivalent lanthanides: *hcp*-(Sm-type)-*dhcp*-*fcc* [2].

Under pressure, the valence of Eu undergoes a continuous transition to a mixed-valence state, and $L_{III}$ x-ray absorption studies indicate that the valence increases to 2.64 at 18 GPa, and is then pressure independent up to 33 GPa [3]. Very recently, Eu has been found to be superconducting at a pressure of approximately 80 GPa, with $T_c = 1.8$ K [4].

Initial high-pressure x-ray diffraction studies by Takemura and Syassen found that Eu exhibits the same body-centred cubic (*bcc*) structure as the divalent alkaline metals at ambient pressure, which transforms to a hexagonal close-packed (*hcp*) structure at 12.5 GPa and then to a new phase (Eu-III) at 18 GPa [1]. Eu-III was initially and tentatively assigned a hexagonal structure closely related to a superstructure of *hcp*.

Krüger *et al.* subsequently performed energy-dispersive x-ray diffraction up to 40 GPa [5]. At pressures exceeding 32 GPa, they found the diffraction pattern to become more complex, with the appearance of new reflections in addition to those observed in the 18–32 GPa region. It was noted that the additional reflections were weak in comparison to the *hcp* peaks, and it was suggested that they could be attributed to superlattice distortions of the *hcp* lattice, or to phase mixing.

Despite this interesting behaviour, very little further attention has been paid to the high-pressure behaviour of Eu until this year, when, prompted by the observation of superconductivity, Bi *et al.* reported results from an x-ray diffraction study up to 92 GPa [6]. They attributed the complex diffraction patterns observed above 18 GPa to arise from a sluggish transition to a mixture of *hcp* and a monoclinic phase with space group *C2/c*. Bi *et al.* did not observe a transition at 33 GPa, but at pressures above 41 GPa, changes in the diffraction pattern were attributed to a mixture of the *C2/c* phase and an orthorhombic phase with the space group *Pnma*, followed by a transition to a pure *Pnma* phase at pressures exceeding 66 GPa. These diffraction results were supported by *ab initio* structure prediction calculations.

In this paper, we present angle-dispersive x-ray diffraction data to illustrate that the Eu-III phase does not consist of pure Eu, but in fact consists of *hcp*-Eu plus a rhombohedral phase, the atomic volume of which suggests it is an impurity.

## 2. Experimental

Angle-dispersive x-ray diffraction was performed at the SRS and ESRF synchrotron sources using monochromatic x-rays of wavelengths 0.44397 Å and 0.4161 Å, respectively. The x-ray beams were collimated to diameters of 50 μm and 15 μm, and the data were collected using MAR345 and MAR555 detectors, respectively.

High-purity Eu samples, supplied by U. Schwarz at the Max-Planck-Institut für Chemische Physik fester Stoffe in Dresden, were loaded into diamond-anvil pressure cells in a dry oxygen-free environment (<1 ppm $O_2$ and <1 ppm $H_2O$) using rhenium and tungsten gaskets. A small piece of ruby was included for pressure calibration. Samples were loaded using helium, mineral oil and no pressure-transmitting media (PTM). The diffraction patterns were integrated using Fit2D and analysed using Rietveld and Le Bail methods with the JANA2006 software [7-9].

## 3. Discussion

In our initial samples, including those loaded using a PTM (helium, mineral oil) and those loaded without a PTM, we observed the appearance of weak non-*hcp* reflections at 18 GPa, in agreement with all previous studies [1,5,6]. In these samples we found that as the pressure is increased further, the intensity ratio between the strong *hcp* peaks and the weak extra peaks remains constant all the way to a phase transition at 33 GPa. This behaviour was also noted by Takemura and Syassen [1]. However, we found that the relative intensity of the *hcp* and extra peaks varies between different samples. Additionally, we left one sample of Eu in mineral oil at 26.1 GPa for one month in order to investigate changes over time. The intensity of the non-*hcp* reflections grew significantly, and many further peaks could be identified. In addition, we found that the Debye-Scherrer rings on the diffraction image corresponding to the non-*hcp* reflections had become spotty, while those from the *hcp* phase remained smooth. This is illustrated in figure 1. All of this behaviour strongly indicated that there were two phases present, and we therefore attempted to index the non-*hcp* peaks as a separate phase.

The extra phase was indexed as having a rhombohedral unit cell with $a = 9.293(4)$ Å, and $c = 5.381(4)$ Å at 26.1 GPa. The lattice parameters of the *hcp* phase at the same pressure are $a = 3.159(1)$ Å and $c = 4.869(1)$ Å. Analysis of the systematic absences of the rhombohedral phase showed the spacegroup to be $R\bar{3}c$ or $R3c$, while density considerations restricted the number of Eu atoms per hexagonal unit cell to 18. Trial refinements showed an excellent fit could be obtained with atoms occupying the 18e ($x$,0,¼) Wyckoff sites of $R\bar{3}c$, with $x = 0.800(4)$. An identical structure was obtained in spacegroup *R3c*. A two-phase *hcp*/ $R\bar{3}c$ Rietveld refinement at 26.1 GPa is shown in figure 2.

The volume per Eu atom of the *hcp* phase at 26.1 GPa is 21.05(1) Å$^3$, while the $R\bar{3}c$ phase at the same pressure is *less* dense, with a volume/atom of 22.36(1) Å$^3$. The decrease in density of 5.86(6) % suggests that the $R\bar{3}c$ phase is not pure Eu, but rather results from a pressure-induced reaction, and due to the small difference in the vol/atom, is perhaps a hydride. The quality of the Rietveld fit shown in figure 2 is excellent, suggesting that whatever the contaminant atoms are, they are weakly-scattering in comparison with europium atoms.

In order to confirm that the $R\bar{3}c$ phase was not pure Eu, we loaded samples with no pressure-transmitting medium and no ruby spheres in a very-high quality glovebox (<0.1 ppm $O_2$ and <0.1 ppm

H$_2$O) in order to obtain contaminant-free samples. In such samples, we observed no additional peaks appearing at 18 GPa, but only a single-phase *hcp* pattern from 12.5 – 33 GPa (see figure 3). However, heating such a sample for 2.5 hours at 100 °C resulted in the reappearance of the spotty Debye-Scherrer rings from the *R–3c* phase.

Peaks corresponding to the Eu-III impurity phase have been observed in all previous x-ray diffraction studies performed to pressures above 18 GPa [1,5,6], and we observed the same peaks in all of our samples loaded with a pressure medium, or pressure calibrant, or loaded in a non-optimum glovebox (>1 ppm O$_2$ and >1 ppm H$_2$O). Indeed, diffraction patterns shown in Ref [5] from the *hcp* phase at 14 GPa show evidence of a contaminant phase at this lower pressure. This illustrates the extreme reactivity of Eu, and the difficulty in loading clean, contaminant-free samples. Our experience suggests that it is more difficult to load contaminant-free Eu samples than it is to load contaminant-free Rb and Cs. The present study also suggests that future studies investigating the pressure-induced superconductivity or valence change of Eu should be combined with x-ray diffraction studies of the same samples to ensure that they are free from contaminant phases.

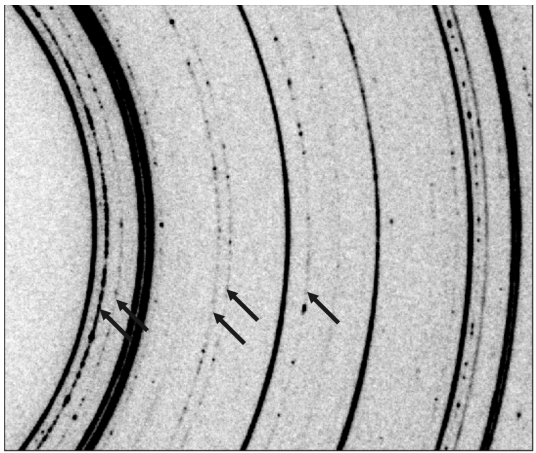

**Figure 1.** A 2D diffraction image from Eu at 26.1 GPa. The Debye-Scherrer (D-S) rings corresponding to the non-*hcp* reflections are spotty, whereas the *hcp* D-S rings are smooth. The arrows mark the five additional reflections observed by Takemura and Syassen [1].

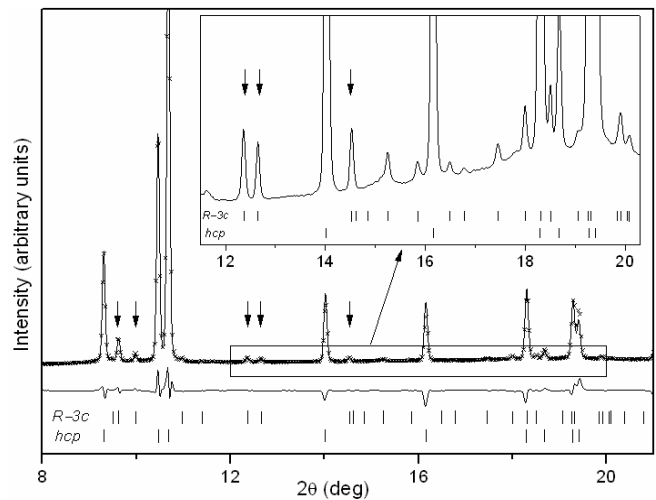

**Figure 2.** A two-phase *hcp*/ $R\bar{3}c$ Rietveld refinement of Eu-III at 26.1 GPa. The experimental data are shown by the crosses and the fit by the solid line. The residuals are shown under the fit and the tick marks show the calculated peak positions. The arrows indicate the Eu-III reflections observed by Takemura and Syassen [1]. The inset shows an enlarged view of the high-angle part of the profile.

Using a contaminant-free sample is also vital for understanding the structural behaviour of europium above 33 GPa, in order to determine which diffraction peaks result from pure-Eu alone. Increasing the pressure of our contaminant-free sample to 33 GPa, we observed the appearance of many additional weak peaks in agreement with Krüger *et al.* [5]. Our data also clearly show splittings of the *hcp* peaks which have not been resolved in previous studies [1,5]. These splittings suggest that the new phase has a distorted-*hcp* structure, and we initially considered the orthorhombic *Pnma* structure proposed by Bi *et al.* for pressures exceeding 41 GPa. A refinement of this structure using the Le Bail method to a profile collected from Eu at 34 GPa is shown in figure 4. The best-fitting lattice parameters are $a$ = 5.291(1) Å, $b$ = 4.713(1) Å, and $c$ = 3.085(1) Å. Although this structure is an orthorhombic distortion of *hcp*, and results in a splitting of the *hcp* diffraction peaks, it cannot account for the observed splittings, and the overall fit is very poor, as illustrated by the inset. There are also a large number of extra peaks not accounted for by this structure. More complex structures must therefore be considered to obtain the correct solution.

In conclusion, we have made high-resolution powder diffraction studies of europium to 37 GPa. These reveal that the long-reported phase transition at 18 GPa to the Eu-III phase is, in fact, a pressure-induced reaction, resulting in the sample becoming a mixture of *hcp*-Eu and a rhombohedral contaminant, perhaps a hydride. Very careful sample preparation and pressure cell loading is required to obtain samples without the contaminant phase. The reasons for the reaction to occur at 18 GPa are unclear, but may be related to the valence of Eu become close to 2.666 at this pressure, perhaps aiding the creation of a stochiometric compound.

**Acknowledgements**
We thank the European Synchrotron Radiation Facility (ESRF) and Daresbury Laboratory (SRS) for provision of synchrotron time and support. We thank M. Hanfland of ID09a (ESRF) and A. Lennie of 9.5HPT (SRS) for their help with the experiments. This work was supported by a research grant and a fellowship (I.L.) from the UK Engineering and Physical Sciences Research Council.

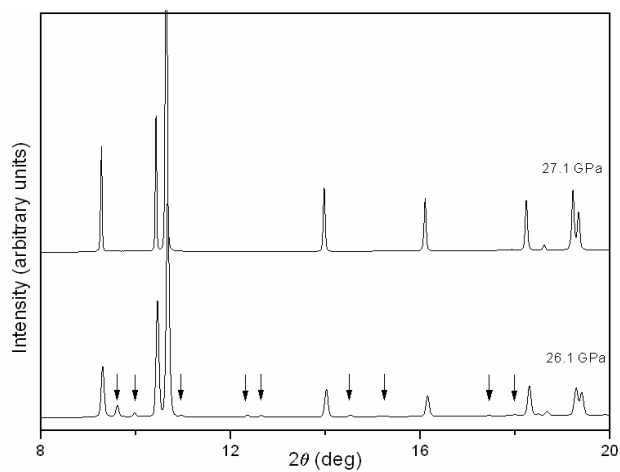
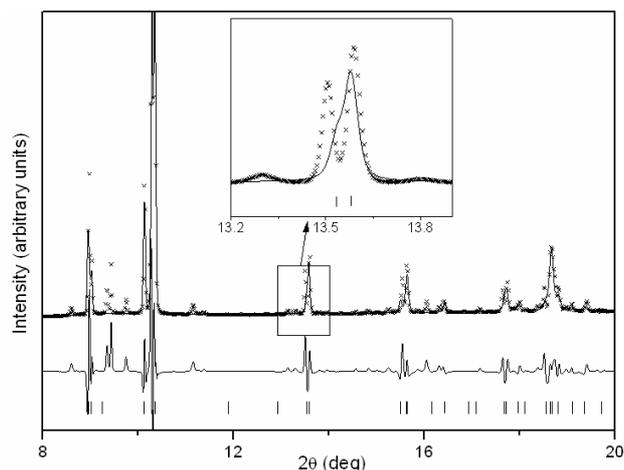

**Figure 3.** Integrated diffraction patterns from a contaminant-free sample of Eu at 27.1 GPa (upper) and a contaminated sample of Eu at 26.1 GPa (lower). Arrows in the lower profile mark the positions of the most intense $R\bar{3}c$ peaks.

**Figure 4.** Le Bail refinement of a diffraction profile from a contaminant-free sample of Eu at 34 GPa, using the *Pnma* structure proposed by Bi *et al.* [5] for Eu above 41 GPa. The residuals are shown under the fit and the tick marks show the calculated peak positions. The inset shows an enlarged view of the fit to the split (102) peak from the *hcp* phase.